\documentclass[aps,prl,showpacs,amsmath,amssymb,superscriptaddress,reprint,10pt]{revtex4-1}
\usepackage{bm}
\usepackage{graphicx}
\usepackage{xcolor}
\usepackage{url}
\usepackage[normalem]{ulem}
\usepackage{times}
\usepackage{bbm}
\usepackage{mathrsfs} 

\usepackage[colorlinks=true,breaklinks=true,allcolors=blue]{hyperref}
\usepackage{url}

\definecolor{johannes}{rgb}{.5,1,.5}

\definecolor{lorenzo}{rgb}{1,.4,.3}

\definecolor{analabha}{rgb}{255, 0, 255}

\definecolor{michael}{rgb}{0,.8,.5}

\newcommand\CC{{\mathbbm{C}}}
\newcommand\id{{\mathbbm{1}}}

\newcommand\dd{{\mathrm{d}}}
\newcommand\ee{{\mathrm{e}}}
\newcommand\ii{{\mathrm{i}}}
\newcommand{\mvec}[1]{\boldsymbol #1}
\newcommand{\Com}[2]{\left[{#1},{#2}\right]}

\DeclareMathOperator{\Tr}{{Tr}}

\allowdisplaybreaks

\begin{document}


\title{Simulation of Quantum Spin Dynamics by Phase Space Sampling of BBGKY Trajectories}  

\author{Lorenzo Pucci} 
\affiliation{National Institute for Theoretical Physics (NITheP), Stellenbosch 7600, South Africa} 

\author{Analabha Roy} 
\affiliation{National Institute for Theoretical Physics (NITheP), Stellenbosch 7600, South Africa} 

\author{Michael Kastner} 
\email{kastner@sun.ac.za} 
\affiliation{National Institute for Theoretical Physics (NITheP), Stellenbosch 7600, South Africa} 
\affiliation{Institute of Theoretical Physics,  University of Stellenbosch, Stellenbosch 7600, South Africa}

\date{\today}

\begin{abstract}
A numerical method, suitable for the simulation of the time evolution of quantum spin models of arbitrary lattice dimension, is presented. The method combines sampling of the Wigner function with evolution equations obtained from the Bogoliubov-Born-Green-Kirkwood-Yvon (BBGKY) hierarchy. Going to higher orders of the BBGKY hierarchy allows for a systematic refinement of the method. Quantum correlations are treated through both, the Wigner function sampling and the BBGKY evolution, bringing about highly accurate estimates of correlation functions. The method is particularly suitable for long-range interacting systems, and we demonstrate its power by comparing with exact results as well as other numerical methods. As an application we compute spin squeezing in a two-dimensional lattice with power-law interactions and a transverse field, which should be accessible in future ion trap experiments.
\end{abstract}


\maketitle 

With the development of highly controllable platforms for quantum simulation, experimental realizations of a variety of, among others, spin lattice models have become available \cite{Simon_etal11,*Struck_etal11,*Labuhn,Britton_etal12,Schauss_etal12,*Islam_etal13,*dePaz_etal13,*Yan_etal13,*Jurcevic_etal14,*Richerme_etal14}. Many of these realizations allow for the preparation of initial states far from equilibrium, and  subsequent time evolution according to a given spin Hamiltonian. For one-dimensional lattices, reliable numerical methods, often based on matrix product states, are available to complement experimental efforts, at least up to intermediate time scales \cite{Schollwoeck11}. However, in two- and higher-dimensional lattices, computational methods for strongly interacting quantum systems out of equilibrium are scarce, and further development of simulation methods is much needed.

In this Letter we develop a novel simulation method by combining phase space sampling of the initial state with a systematic semi-classical expansion based on the BBGKY hierarchy. The sampling scheme is inspired by a discrete Wigner representation for spin systems as introduced by Wootters \cite{Wootters87}, and was recently used in the context of dynamical simulations by Schachenmayer {\em et al.} \cite{Schachenmayer_etal15,Schachenmayer_etalNJP15}. However, that scheme only incorporates quantum fluctuations on the level of the initial state and thus accounts for them only for short times. We combine the sampling with a systematic way of deriving time evolution equations for arbitrary $n$-point correlations, based on the BBGKY hierarchy \cite{Bonitz}. In the past decade or so, BBGKY techniques have been successfully applied to classical systems with long-range interactions \cite{BouchetDauxois05,*Nardini_etal12}, where the error made when truncating the infinite hierarchy of equations 
is known to vanish with increasing system size \cite{BraunHepp77}. For this reason we expect the phase space sampling method for BBGKY trajectories to be particularly suitable for many-particle systems with long-range interactions. Benchmarking against exactly solvable models indeed confirms that correlations can be computed to a significantly greater accuracy, and indicates that our method is a candidate for the most accurate dynamical simulation technique available for higher-dimensional spin models with long-range interactions.

{\em Discrete Wigner representation.---}For concreteness we use Wootters's representation of a spin-$1/2$ degree of freedom, but higher spin quantum numbers can be treated along similar lines \cite{Wootters87}. Starting point is a discrete phase space $\Gamma=\left\{(0,0),(0,1),(1,0),(1,1)\right\}$ consisting of four points, to each of which a 3-vector is associated, $\mvec{r}_{(0,0)}=(1,1,1)$, $\mvec{r}_{(0,1)}=(-1,-1,1)$, $\mvec{r}_{(1,0)}=(1,-1,-1)$, and $\mvec{r}_{(1,1)}=(-1,1,-1)$. To each phase space point $\alpha\in\Gamma$ one assigns a so-called phase point operator
\begin{equation}\label{e:Aalpha}
A_\alpha = \tfrac{1}{2}(\id+\mvec{r}_\alpha\cdot\mvec{\sigma}),
\end{equation}
where $\mvec{\sigma}=\left(\sigma^x,\sigma^y,\sigma^z\right)$ is the vector of Pauli operators. A density operator $\rho$ on $\CC^2$ can then be written  as a linear combination of phase point operators,
\begin{equation}
\rho=\sum_{\alpha\in\Gamma}w_\alpha A_\alpha,
\end{equation}
where the weights $w_\alpha=\Tr(\rho A_\alpha)/2$ form a quasi-probability distribution analogous to the Wigner function for continuous degrees of freedom \cite{Wootters87}. For $N$ spin-$1/2$ degrees of freedom and assuming a product (initial) state, such a phase space representation generalizes to
\begin{equation}
\rho_0=\sum_{\alpha_1,\dotsc,\alpha_N\in\Gamma}w_{\alpha_1}\cdots w_{\alpha_N}A_{\alpha_1}\otimes\cdots\otimes A_{\alpha_N}.
\end{equation}
If the weights $w_{\alpha_i}$ calculated from $\rho_0$ happen to be all nonnegative, we can sample from this initial state by drawing $N$-spin phase space points $\mvec{\alpha}=(\alpha_1,\dotsc,\alpha_N)\in\Gamma^N$ according to the probability distribution $w_{\alpha_1}\cdots w_{\alpha_N}$. In practice we use a slightly different, and superior, sampling scheme, which is detailed in \cite{Note1}. This sampling, which is inspired by Ref.~\cite{Schachenmayer_etal15}, is the first main ingredient of our numerical scheme.

{\em BBGKY evolution equations.---}As a second ingredient we derive semi-classical evolution equations with which to propagate the sampled initial phase space points. To this purpose we write the time-evolution of the density operator as
\begin{equation}\label{e:rho_t}
\rho(t)=\sum_{\alpha_1,\dotsc,\alpha_N\in\Gamma}w_{\alpha_1}\cdots w_{\alpha_N} \mathscr{A}_{1\dotsc N}^{\alpha_1\dotsc\alpha_N}(t),
\end{equation}
where the operators $\mathscr{A}_{1\dotsc N}^{\alpha_1\dotsc\alpha_N}(t)=\ee^{-\ii Ht}A_{\alpha_i}\otimes\cdots\otimes A_{\alpha_N}\ee^{\ii Ht}$  have unit trace and satisfy a Liouville-von Neumann equation,
\begin{equation}\label{VNeqdWA}
\ii\partial_t \mathscr{A}_{1\dotsc N}^{\alpha_1\dotsc\alpha_N} = \Com{H}{\mathscr{A}_{1\dotsc N}^{\alpha_1\dotsc\alpha_N}}.
\end{equation}
Unlike Ref.~\cite{Schachenmayer_etal15}, we chose to write the time evolution \eqref{e:rho_t} in the Schr\"odinger picture, as this naturally leads to the systematic approximation scheme introduced in the following.

While the operators $\mathscr{A}_{1\dotsc N}^{\alpha_1\dotsc\alpha_N}$ are in general not positive definite and therefore do not qualify as density operators, reduced $\mathscr{A}$-operators (analogous to reduced density operators) can be defined by tracing out parts of the system,
\begin{equation}
\mathscr{A}_i^{\alpha_1\dotsc\alpha_N}=\Tr_{\not{\,i}} \mathscr{A}_{1\dotsc N}^{\alpha_1\dotsc\alpha_N},\quad \mathscr{A}_{ij}^{\alpha_1\dotsc\alpha_N}=\Tr_{\not{\,i}\not{\,j}} \mathscr{A}_{1\dotsc N}^{\alpha_1\dotsc\alpha_N},
\end{equation}
where $i\neq j$. Here, $\Tr_{\not{\,i}}$ denotes a partial trace over all of the tensor product Hilbert space except for the factor associated with spin $i$. In the spirit of the BBGKY hierarchy for reduced density operators \cite{Bonitz}, the time evolution \eqref{VNeqdWA} induced by a general Hamiltonian
\begin{equation}\label{e:Hgen}
H_{1\dotsc N} = \sum_i H_i + \sum_{i<j}H_{ij}
\end{equation}
with on-site and pair interactions can be recast in the form of a hierarchy of equations for the $n$-spin reduced $\mathscr{A}$-operators \cite{Note1}. For the long-range interacting systems we intend to simulate, we expect deviations from mean-field to be small. This suggests to separate the $\mathscr{A}$-operators into product and correlated parts by means of a cluster expansion,
\begin{subequations}
\begin{align}
\mathscr{A}_{ij}&=\mathscr{A}_i \mathscr{A}_j+\mathscr{C}_{ij},\label{e:cluster1}\\
\mathscr{A}_{ijk}&=\mathscr{A}_i \mathscr{A}_j \mathscr{A}_k + \mathscr{A}_i \mathscr{C}_{jk} + \mathscr{A}_j \mathscr{C}_{ik} + \mathscr{A}_k \mathscr{C}_{ij} + \mathscr{C}_{ijk},\label{e:cluster2}
\end{align}
\end{subequations}
and similarly for higher orders. Superscripts $\alpha_1\dotsc\alpha_N$ of $\mathscr{A}$ and $\mathscr{C}$ are for the moment suppressed. In terms of these operators the first two equations of the BBGKY hierarchy read
\begin{subequations}
\begin{align}
\ii\partial_t \mathscr{A}_i=&\Com{H_i}{\mathscr{A}_i}+\sum_{k\neq i}\Tr\Com{H_{ik}}{\mathscr{C}_{ik}+\mathscr{A}_i \mathscr{A}_k}\label{e:1st_order}\\
\ii\partial_t \mathscr{C}_{ij}=&\Com{H_i+H_j+H_{i\not{\,j}}^\text{H}+H_{j\not{\,i}}^\text{H}}{\mathscr{C}_{ij}}+\Com{H_{ij}}{\mathscr{C}_{ij}+\mathscr{A}_i \mathscr{A}_j}\nonumber\\
&+\sum_{k\neq i,j}\left(\Tr_k\Com{H_{ik}}{\mathscr{A}_i \mathscr{C}_{jk}}+\Tr_k\Com{H_{jk}}{\mathscr{A}_j \mathscr{C}_{ik}}\right)\nonumber\\
&-\mathscr{A}_i\Tr_i\Com{H_{ij}}{\mathscr{C}_{ij}+\mathscr{A}_i \mathscr{A}_j}\nonumber\\
&-\mathscr{A}_j\Tr_j\Com{H_{ij}}{\mathscr{C}_{ij}+\mathscr{A}_i \mathscr{A}_j}\label{e:2nd_order}
\end{align}
\end{subequations}
with the Hartree operator
\begin{equation}
 H_{i\not{\,j}}^\text{H}=\sum_{k\neq i,j}\Tr_k\left(H_{ik} \mathscr{A}_k\right).
\end{equation}
In \eqref{e:2nd_order} we have neglected the 3-spin correlations $\mathscr{C}_{ijk}$, and this approximation is expected to be good for intermediate times and sufficiently long-ranged interactions. Improving this scheme systematically by truncating the BBGKY hierarchy at higher order is straightforward, at a numerical cost that scales like $\mathscr{O}\left(N^n\right)$ with system size $N$ and truncation order $n$. If we were to neglect also the 2-spin correlations $\mathscr{C}_{ij}$ in \eqref{e:1st_order}, and disregard \eqref{e:2nd_order} entirely, we would recover the time evolution equations used in Ref.~\cite{Schachenmayer_etal15}.

Next, we represent the $\mathscr{A}$- and $\mathscr{C}$-operators in the basis of Pauli operators,
\begin{subequations}
\begin{align}
\mathscr{A}_i &= \tfrac{1}{2}\left(\id_i+\mvec{a}_i\cdot\mvec{\sigma}_i\right),\label{e:Aexp}\\
\mathscr{C}_{ij} &= \tfrac{1}{4}\sum_{\mu,\nu\in\{x,y,z\}}c_{ij}^{\mu\nu}\sigma_i^\mu\sigma_j^\nu,\label{e:Cexp}
\end{align}
\end{subequations}
for $i\neq j$. Inserting these expansions into the BBGKY equations \eqref{e:1st_order} and \eqref{e:2nd_order}, we obtain time evolution equations for the expansion coefficients $a_i^\mu$ and $c_{ij}^{\mu\nu}$ \cite{PaskauskasKastner12}; see \cite{Note1} for a derivation. This set of $3N(3N-1)/2$ coupled ordinary differential equations is the second main ingredient of our simulation method.

\begin{figure}\centering
\includegraphics[width=\linewidth]{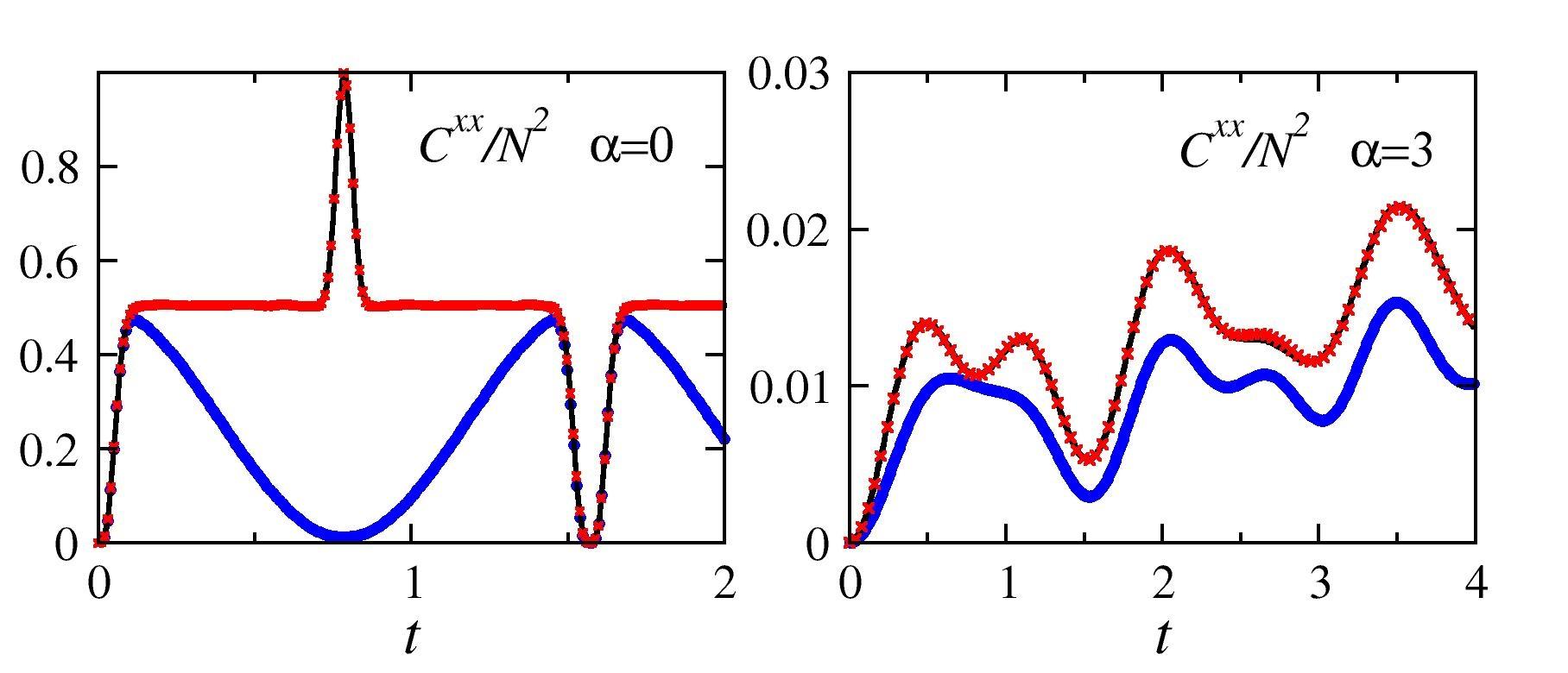}
\caption{\label{f:Ising}%
Time evolution of total connected correlations $C^{xx}$ \eqref{e:Cmumu} for a long-range Ising chain of $100$ sites in the absence of a magnetic field and with long-range exponent $\alpha=0$ (left) and $\alpha=3$ (right). As shown analytically in \cite{Note1}, our method recovers the exact solution (black) in the limit of an infinite sample. The only error (which is not visible on the scale of the plot) is therefore of a statistical nature due to the finite sample size $n=10000$ with $S_{1-1}$ sampling \cite{Note1} (red crosses). The method of SPR (blue dots) shows systematic deviations from the exact result.
}%
\end{figure}

In summary, the simulation method consists of the following steps. (1) Sample phase space points $\mvec{\alpha}$ from the probability distribution $w_{\alpha_1}\cdots w_{\alpha_N}=2^{-N}\Tr\left[\rho_0 \mathscr{A}_{1\dotsc N}^{\alpha_1\dotsc\alpha_N}(0)\right]$ of the initial state $\rho_0$. (2) Compute the corresponding initial values of the Pauli expansion coefficients $a_i^\mu=\Tr\left(\sigma_i^\mu \mathscr{A}_{1\dotsc N}^{\alpha_1\dotsc\alpha_N}\right)$. The correlation coefficients $c_{ij}^{\mu\nu}$ are initially zero for a product state, but the method can be extended to correlated initial states by using nonzero initial values. (3) Time-evolve $a_i^\mu$ and $c_{ij}^{\mu\nu}$ according to the semi-classical equations of motion \eqref{e:1st_order_param}--\eqref{e:2nd_order_param} in \cite{Note1}. (4) Calculate expectation values of 1- or 2-spin functions,
\begin{subequations}
\begin{align}
\left\langle\sigma_i^\mu\right\rangle =& \sum_{\mvec{\alpha}\in\Gamma^N}w_{\alpha_1}\cdots w_{\alpha_N} a_i^\mu \approx \frac{1}{n}\sum_{\mvec{\alpha}\in S_n} a_i^\mu,\label{e:1spin}\\
\left\langle\sigma_i^\mu\sigma_j^\nu\right\rangle =& \sum_{\mvec{\alpha}\in\Gamma^N}w_{\alpha_1}\cdots w_{\alpha_N} \left(c_{ij}^{\mu\nu}+a_i^\mu a_j^\nu\right)\nonumber\\
\approx& \frac{1}{n}\sum_{\mvec{\alpha}\in S_n} \left(c_{ij}^{\mu\nu}+a_i^\mu a_j^\nu\right),\label{e:2spin}
\end{align}
\end{subequations}
for $i\neq j$. Here, $S_n=\left\{\mvec{\alpha}^{(1)},\dots,\mvec{\alpha}^{(n)}\right\}$ is a sample of $n$ phase space points drawn from the probability distribution $w_{\alpha_1}\cdots w_{\alpha_N}$.

{\em Benchmarking.---}To test the accuracy of the simulation method, we benchmark the results against exact analytic and numeric calculations for simple one-dimensional models. All simulations are done for fully $x$-polarized initial states $\rho_0=2^{-N}(\id+\sigma^x)^{\otimes N}$. The improved performance of our method is highlighted by comparing to the results of Schachenmayer, Pikovski, and Rey \cite{Schachenmayer_etal15} (abbreviated to SPR in the following), one of the most powerful methods available for the simulation of higher-dimensional spin models.

The first test case is the quantum Ising chain with long-range interactions,
\begin{equation}\label{e:Ising}
H=-J\sum_{i\neq j}\frac{\sigma_i^z\sigma_j^z}{|i-j|^\alpha} - \mvec{h}\cdot\sum_i\mvec{\sigma}_i,
\end{equation}
where $J$ is a coupling constant (which is set to unity in the following) and $\mvec{h}$ the magnetic field vector. The exponent $\alpha\geq0$ controls the range of the interactions, and open boundary conditions are used. In the case of a longitudinal magnetic field $\mvec{h}=(0,0,h)$, exact analytic results are known for the time-evolution of 1-spin \cite{Emch66,*Radin70,*Kastner11,*Kastner12} and 2-spin functions \cite{vdWorm_etal13,*FossFeigHazzardBollingerRey13,*KastnerVdWorm15}. In \cite{Note1} we show that, for this and also more general models, the middle equations in \eqref{e:1spin} and \eqref{e:2spin} agree with the exact analytic solution, and hence the estimates on the right-hand side of those equations become exact in the limit of large sample size. In this sense, our method is numerically exact for the Ising model in a longitudinal field. A comparison with the method of SPR, which show deviations from the exact result, is shown in Fig.~\ref{f:Ising} for the total connected correlations
\begin{equation}\label{e:Cmumu}
C^{\mu\mu}=\sum_{i,j}\left(\left\langle\sigma_i^\mu\sigma_j^\mu\right\rangle - \left\langle\sigma_i^\mu\right\rangle\left\langle\sigma_j^\mu\right\rangle\right)
\end{equation}
(see Sec.~C of \cite{Note1} on details of how total connected correlations are approximated).

\begin{figure}\centering
\includegraphics[width=\linewidth]{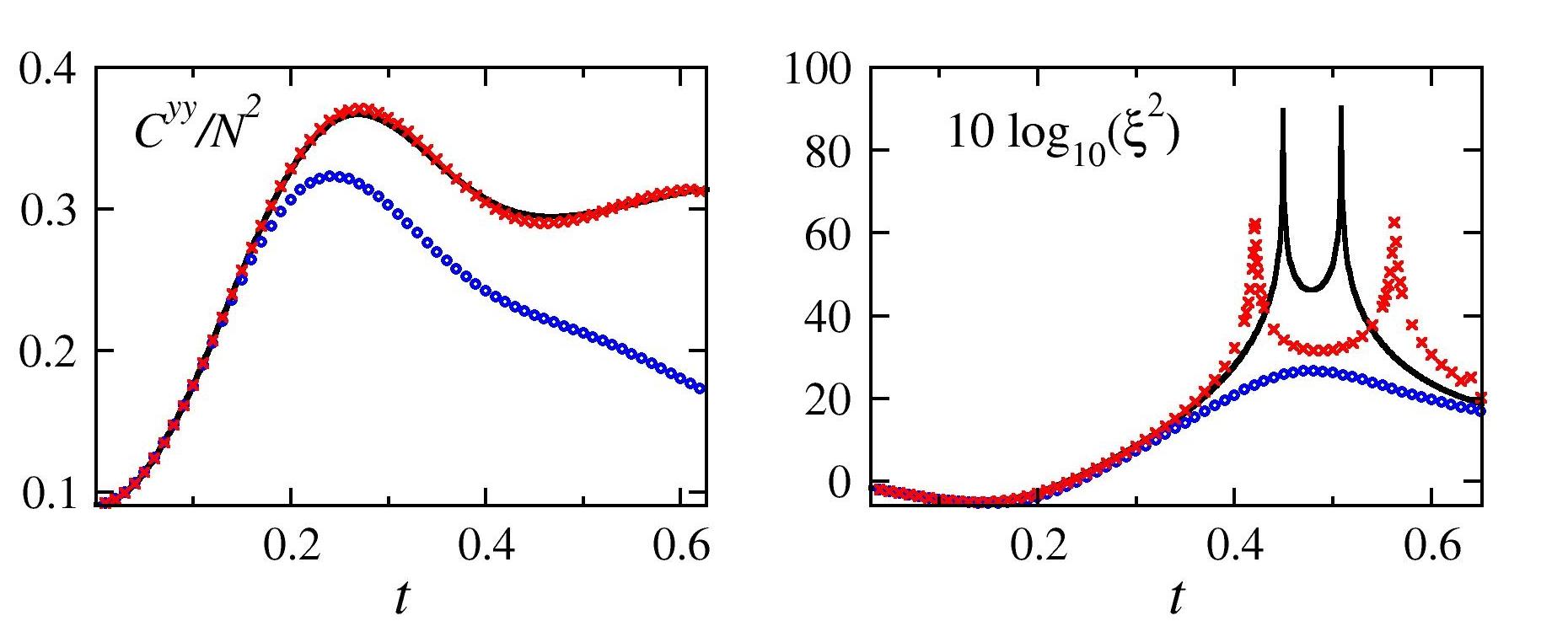}
\caption{\label{f:TFIM}%
Time evolution of a long-range Ising chain of 11 sites with $\alpha=1/2$ in a transverse field of strength $h=1$. We compare exact diagonalization results (black line), the method of SPR (blue dots), and our method with $S_\text{all}$ sampling \cite{Note1} (red crosses) for sample sizes $n=5\times10^5$. Left: Total connected correlations $C^{yy}$ \eqref{e:Cmumu}. Right: Decibel spin squeezing \eqref{e:spinsqueezing}. While not an exact match, our method captures the sharp spikes in the spin squeezing.
}%
\end{figure}

A magnetic field $\mvec{h}=(h,0,0)$ in \eqref{e:Ising} results in a long-range quantum Ising chain in a transverse magnetic field, and no analytic solution is known in this case. We calculated 2-spin correlation functions as well as the spin squeezing parameter~\cite{Wineland_etal92,*Ma_etal11}
\begin{equation}\label{e:spinsqueezing}
\xi = \sqrt{N}\min_{\mvec{n}\perp\langle\mvec{S}\rangle}\frac{\sqrt{\langle(\mvec{S}\cdot\mvec{n})^2\rangle - \langle\mvec{S}\cdot\mvec{n}\rangle^2}}{|\langle\mvec{S}\rangle||\mvec{n}|},
\end{equation}
where $\mvec{S}=\sum_i \mvec{\sigma}_i$. Spin squeezing is an entanglement witness  and can be measured for example in trapped-ion realizations of spin models \cite{Meyer_etal01,*Uys_etal12,*Hazzard_etal14}. A comparison to exact diagonalization results for a chain of 11 sites, and also to the method of SPR, is shown in Fig.~\ref{f:TFIM}. For the chosen parameter values (specified in the figure caption) the spin squeezing shows pronounced spikes at certain times, which are captured by our method. From theoretical arguments we expect the approximation error to become smaller with increasing system size for exponents $\alpha$ smaller than the lattice dimension. 

\begin{figure}\centering
\includegraphics[width=\linewidth]{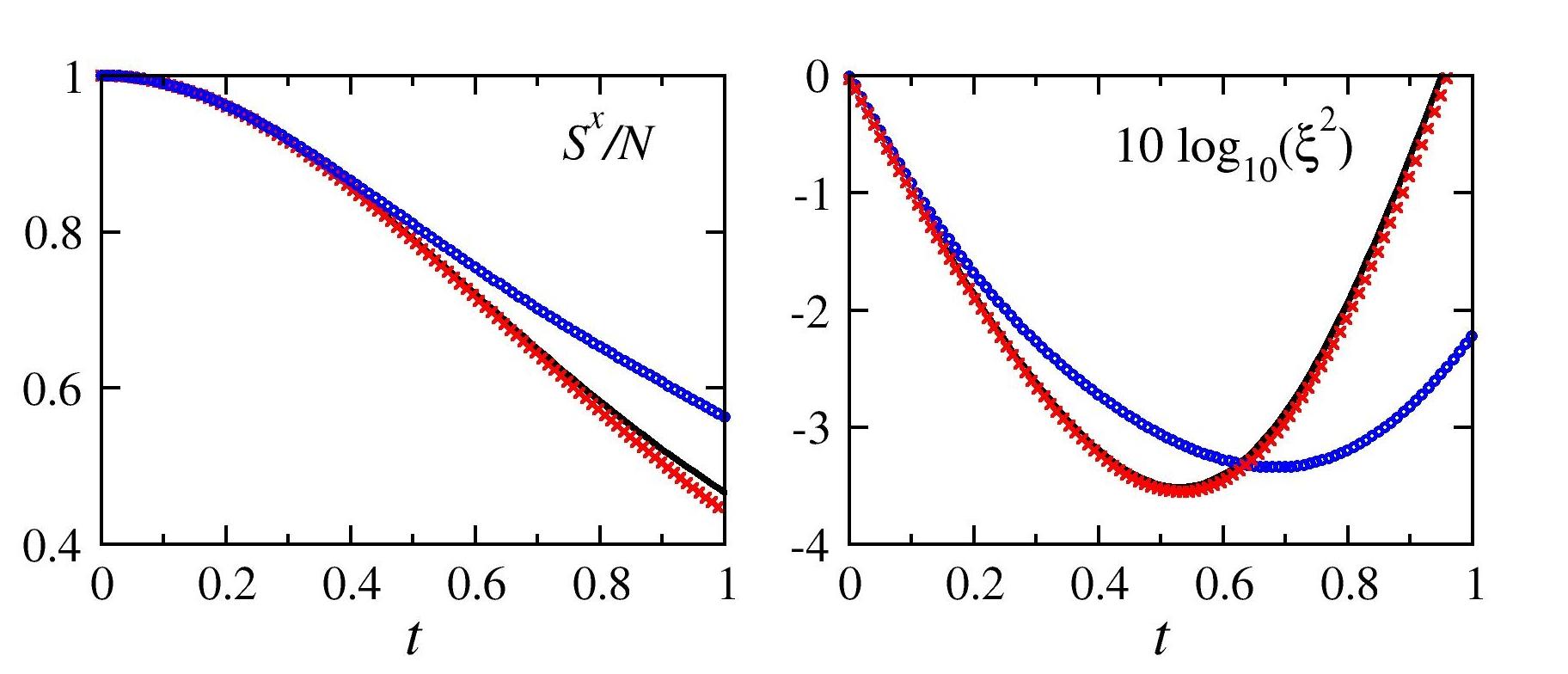}
\caption{\label{f:XX}%
Time evolution of a long-range $XX$ chain of 100 sites with $\alpha=3$. We compare DMRG results (black line), the method of SPR (blue dots), and our method with $S_{0-1}$ sampling \cite{Note1} (red crosses) for sample sizes $n=10^5$. Left: Total spin $S^x=\sum_i\sigma_i^x$. Right: Decibel spin squeezing as obtained from \eqref{e:spinsqueezing}, where the results of our method are virtually indistinguishable from the DMRG data.
}%
\end{figure}

For further benchmarking, we consider the quantum $XX$ chain with long-range interactions,
\begin{equation}\label{e:XY}
H=-J\sum_{i\neq j}\frac{\sigma_i^x\sigma_j^x+\sigma_i^y\sigma_j^y}{|i-j|^\alpha}.
\end{equation}
Here we simulate chains of 100 sites and compare to numerical results from density matrix renormalization group (DMRG) calculations. As shown in Fig.~\ref{f:XX}, correlations and spin squeezing are obtained by our method with remarkable precision.

\begin{figure}\centering
\includegraphics[width=\linewidth]{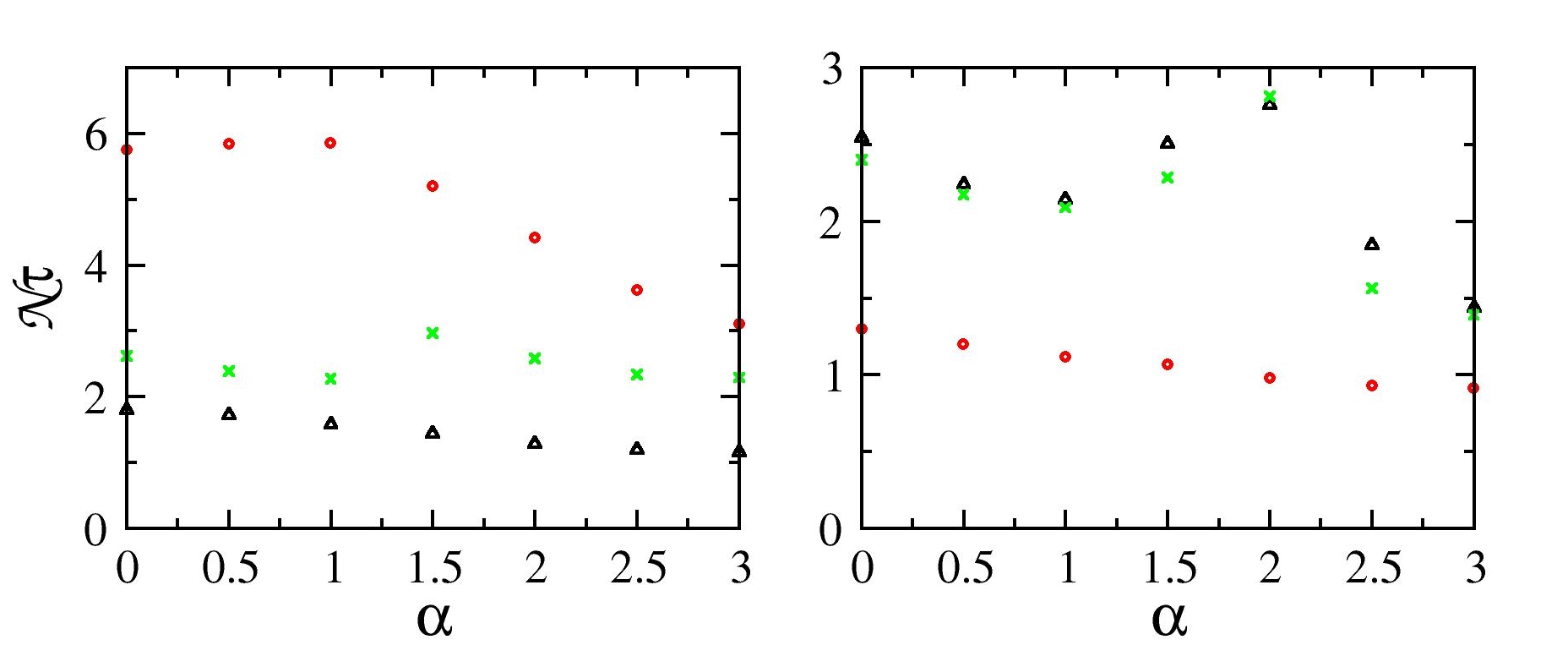}
\caption{\label{f:alpha_dependence}%
Time scale $\tau$ until which the numerically simulated decibel spin squeezing differs from the exact result by less than one. To make the data for different $\alpha$ comparable, $\tau$ is rescaled by the mean energy per lattice site $\mathscr{N}$. The various symbols and colors denote different sampling schemes as defined in \eqref{e:S11}--\eqref{e:Sall}, $S_{\text{1-1}}$ (black), $S_{\text{0-1}}$ (red), and $S_{\text{all}}$ (green). Left: $XX$ chain of length $N=11$ with $J=-1/2$. Right: Ising chain of length $N=11$ in a transverse field $h=1$ with coupling $J=-1$.
}%
\end{figure}

{\em Parameter dependence.---}Based on the theoretical reasoning outlined in the introduction, we expect the simulation method to perform best for systems with long-range interactions. To test this prediction, we studied the accuracy of the simulation results of the decibel spin squeezing for various values of the long-range exponent $\alpha$ by comparing numerical to exact results. We then extracted the time scale $\tau$ up to which the deviation from the exact result is smaller than one. In Fig.~\ref{f:alpha_dependence} we plot $\tau$, rescaled by the mean energy per lattice site $\mathscr{N}=\sum_{i\neq j} (J_{ij}^x+J_{ij}^x+J_{ij}^z)/N$, as a function of $\alpha$. While a trend towards better performance for small $\alpha$ is visible, the $\alpha$ dependence is rather weak, indicating that the method performs robustly over a broad range of $\alpha$-values. 

\begin{figure}\centering
\includegraphics[height=0.4\linewidth]{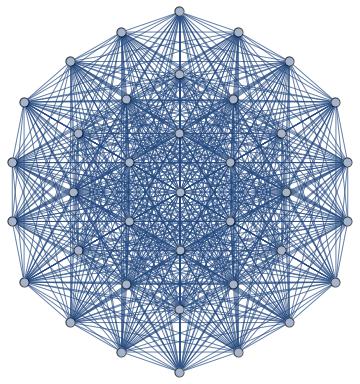}
\hfill
\includegraphics[height=0.4\linewidth]{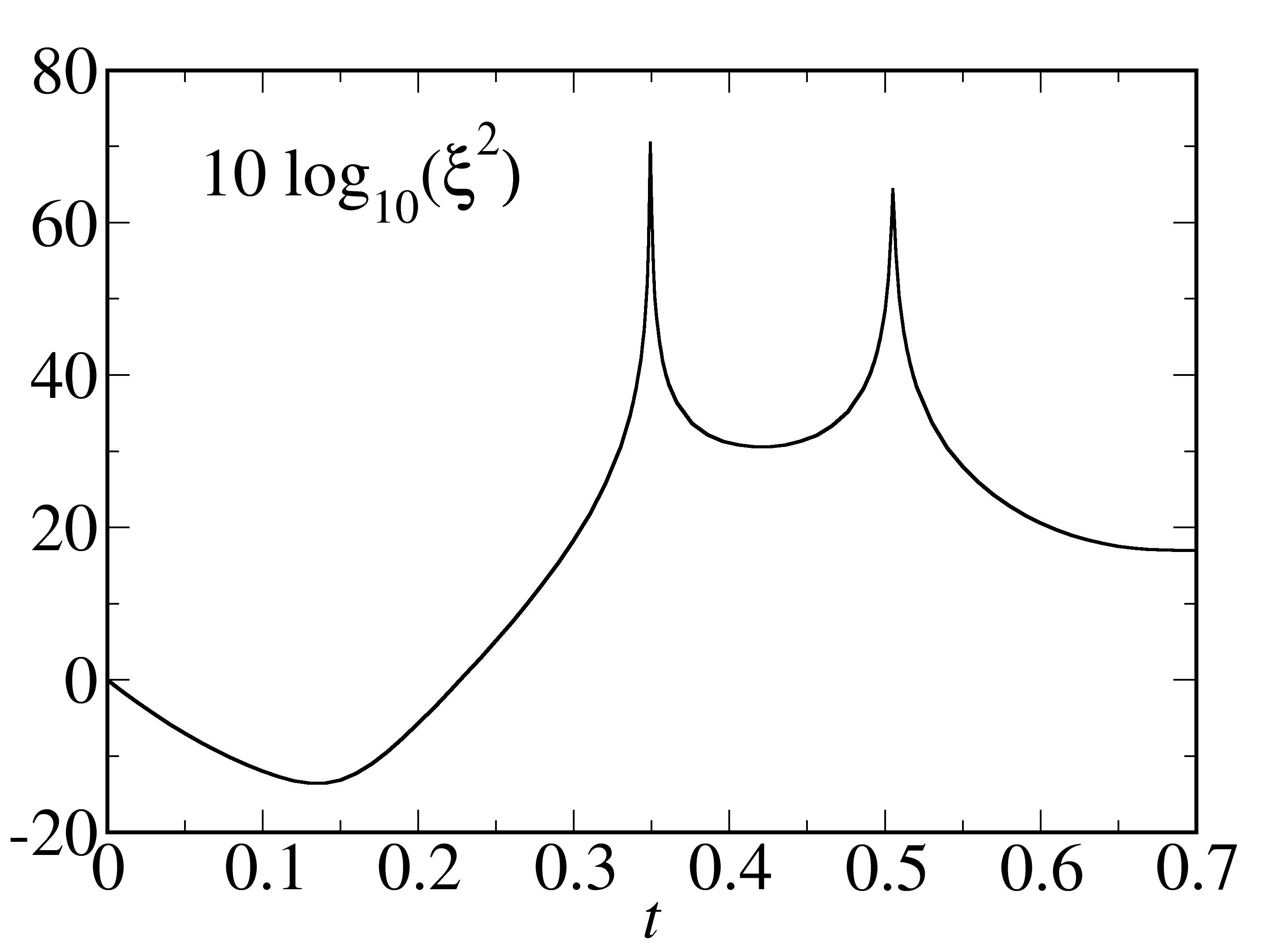}
\caption{\label{f:2d}%
Left: Graph representing a two-dimensional crystal consisting of $N=217$ ions, as it forms spontaneously in the external trapping potential of \cite{Britton_etal12}. The crystal structure is approximately triangular, with deformations close to the boundary. Lattice spacings are of the order of $15\mathrm{\mu m}$. The couplings $J_{ij}$, as indicated by the graph edges, are nonvanishing between all pairs $i$, $j$ of ions, decaying approximately like $J_{ij}\propto|i-j|^{0.7}$. Right: Time evolution of the decibel spin squeezing \eqref{e:spinsqueezing} for an Ising model on such a graph, with couplings $J_{ij}$ and a transverse magnetic field of strength $h=1$. For the numerical calculation we used the $S_\text{all}$ sampling and $n=4\times10^4$ initial states. 
}%
\end{figure}

{\em Application to two-dimensional ion crystals.---}While one-dimensional models are suitable for benchmarking our method, it is two- and higher-dimensional systems, where DMRG calculations are not feasible, for which our method is particularly advantageous. We illustrate its strength by calculating spin squeezing in a transverse-field Ising model on a two-dimensional lattice with hundreds of sites, as it can be emulated with ions in a Penning trap \cite{Britton_etal12}. The Coulomb repulsion between the ions leads approximately to a triangular lattice structure, although distortions occur close to the boundary [see Fig.~\ref{f:2d} (left)]. Transverse lattice oscillations mediate interactions between hyperfine states of the ions, leading to effective Ising spins with long-range couplings, like in a two-dimensional version of the Hamiltonian \eqref{e:Ising}. The effective magnetic field $\mvec{h}$ in the experiment in \cite{Britton_etal12} can be orientated arbitrarily at least in principle, although up 
to now only data for vanishing fields have been obtained. For a longitudinal field $\mvec{h}=(0,0,h)$ exact results are available, 
and are used for benchmarking in the case of vanishing $h$. Here we consider the long-range Ising model with $N=217$ lattice sites in a transverse field $\mvec{h}=(h,0,0)$, where no other good methods are available for calculating how spin squeezing evolves in time. Experiments for this case are planned, and are expected to take place in the coming months or years. The coupling coefficients $J_{ij}$ are computed numerically by a method outlined in \cite{Britton_etal12}. They decay approximately like $J_{ij}\propto|i-j|^{0.7}$ for the experimentally realistic parameters chosen. Numerical results for decibel spin squeezing are shown in Fig.~\ref{f:2d} (right), predicting strong spin squeezing at times around around $t\approx0.15$, as well as spikes of large positive squeezing parameter at later times. These results, and similar ones for other parameter values, can serve as a useful reference for future experiments in two-dimensional ion crystals.

{\em Conclusions.---}We have introduced a numerical method for simulating the time evolution of quantum spin models. The method uses sampling from a discrete phase space representation of the initial state, as recently introduced in \cite{Schachenmayer_etal15}. This ingredient is combined with the time evolution equations \eqref{e:1st_order_param}--\eqref{e:2nd_order_param} obtained from the BBGKY hierarchy, which explicitly account for the time evolution of connected correlations. Benchmarking against exact results confirms that this leads to a significantly improved accuracy when computing correlation functions, as illustrated in Figs.~\ref{f:Ising}--\ref{f:XX}. The numerical cost scales like $\mathscr{O}(N^2)$ with the lattice size $N$. Lattices with $N=\mathscr{O}(10^2)$ are manageable on a personal computer, and $N=\mathscr{O}(10^3)$ can be simulated using distributed high performance computing. Going to higher orders of the BBGKY hierarchy allows for a systematic refinement of the method, but 
results in a less favorable 
scaling of the numerical cost. While 
the method is presented for spin-$1/2$ degrees of freedom, both, the phase space sampling and the BBGKY time evolution equations can be readily generalized to higher spin quantum numbers.

The method is particularly suitable for long-range interacting systems and can be applied in arbitrary lattice dimension. Of particular interest are applications for two- and higher-dimensional lattices, where established methods like DMRG do not work. An important open question concerns the various sampling schemes discussed in \cite{Note1}. These schemes affect the accuracy of the numerics, but a better understanding is required in order to be able to choose the optimal scheme for a given problem.

The authors acknowledge helpful discussions with Lea Santos and Johannes Schachenmayer. John Bollinger and Joe Britton are gratefully acknowledged for providing the coupling coefficients used in the two-dimensional simulations, and Johannes Schachenmayer for the DMRG data used in Fig.~\ref{f:XX}.
A.\,R.\ acknowledges the computational resources provided by the University of Stellenbosch's Rhasatsha HPC, 
and the Tsessebe cluster at the Centre for High Performance Computing, Cape Town.  
M.\,K.\ acknowledges financial support by the National Research Foundation of South Africa via the Incentive Funding and the Competitive Programme for Rated Researchers.

\bibliography{LRLR}

\bigskip

\newpage

\begin{center}
{\bf Supplemental Material}  
\end{center}
\vspace{-3mm}
\appendix
\numberwithin{equation}{section}
\numberwithin{figure}{section}

\section{A. Phase space sampling schemes}
\setcounter{section}{1}
\setcounter{equation}{0}
\setcounter{figure}{0}

In the main text of this Letter, a phase space sampling scheme based upon Wootters's discrete phase space representation is described, where the phase point operators \eqref{e:Aalpha} are defined with the 3-vectors
\begin{subequations}
\begin{align}
\mvec{r}_{(0,0)}&=(1,1,1),\label{e:r1}\\
\mvec{r}_{(0,1)}&=(-1,-1,1),\\
\mvec{r}_{(1,0)}&=(1,-1,-1),\\
\mvec{r}_{(1,1)}&=(-1,1,-1).\label{e:r4}
\end{align}
\end{subequations}
However, this choice is not unique. Phase space operators $A_\alpha^\prime=UA_\alpha U^\dagger$ related to the $A_\alpha$ by an overall unitary transform $U$ also have the desired properties of a Wigner representation, and the same holds for phase point operators obtained by a nonsingular linear transformation of the phase space coordinates \cite{Wootters87}. A simple example is provided by the phase point operators
\begin{equation}
A_\alpha^\prime = \tfrac{1}{2}(\id+\mvec{r}_\alpha^\prime\cdot\mvec{\sigma})
\end{equation}
where the 3-vectors
\begin{subequations}
\begin{align}
\mvec{r}_{(0,0)}^\prime&=(1,-1,1),\\
\mvec{r}_{(0,1)}^\prime&=(-1,1,1),\\
\mvec{r}_{(1,0)}^\prime&=(1,1,-1),\\
\mvec{r}_{(1,1)}^\prime&=(-1,-1,-1),
\end{align}
\end{subequations}
are obtained by flipping the sign of the second component in each of the vectors in \eqref{e:r1}--\eqref{e:r4}.

While both these possibilities, as well as the many others, provide valid discrete Wigner representations, the choice of the phase point operators may significantly affect the phase space sampling discussed in the main text. As an example, consider an initial state $\rho_0 = (\id+\sigma^x)/2$ being a $\sigma^x$-eigenstate with eigenvalue $+1$. Using \eqref{e:r1}--\eqref{e:r4} to compute the phase point operators \eqref{e:Aalpha}, one obtains $w_{(0,0)}=w_{(1,0)}=1/2$ and 
$w_{(0,1)}=w_{(1,1)}=0$. Accordingly, in our simulation method one would time-evolve, each with probability $1/2$, either the classical spin vector $a=(1,1,1)$ or $a=(1,-1,-1)$. The $y$- and $z$-components of these two classical spins vectors are fully correlated, but from a physics perspective there is no good reason for them to be so. It would appear more natural to sample from the four classical spin vectors $(1,1,1)$, $(1,1,-1)$, $(1,-1,1)$, and $(1,-1,-1)$, which combines the vectors $\mvec{r}_{(0,0)}$ and $\mvec{r}_{(1,0)}$ from the original set of 3-vectors with $\mvec{r}_{(0,0)}^\prime$ and $\mvec{r}_{(1,0)}^\prime$ from the set of primed ones. Indeed, this is the kind of sampling that was used in \cite{Schachenmayer_etal15}, and it leads to a significant improvement of the numerical results compared to sampling from $\mvec{r}_{(0,0)}$ and $\mvec{r}_{(1,0)}$ only.

Such a sampling from two different phase space representations may appear as going beyond Wootters's discrete Wigner representation, but the validity of this, and many other, generalized sampling schemes can be understood as follows. Splitting the density operator into two parts, $\rho_0=\rho_0/2+\rho_0/2$, we can choose to write the first term in one kind of Wigner representation, and the second in another one,
\begin{equation}\label{e:split1}
\rho_0=\tfrac{1}{2}\sum_{\alpha\in\Gamma}w_\alpha A_\alpha + \tfrac{1}{2}\sum_{\alpha\in\Gamma}w_\alpha^\prime A_\alpha^\prime.
\end{equation} 
For the initial state $\rho_0 = (\id+\sigma^x)/2$, sampling the spin vectors $\mvec{r}_{(0,0)}$, $\mvec{r}_{(1,0)}$, $\mvec{r}_{(0,0)}^\prime$, and $\mvec{r}_{(1,0)}^\prime$ with probabilities of $1/4$ will therefore give a correct sampling.

It is interesting to note that, since $w_\alpha=w_\alpha^\prime$ for all $\alpha$, we can rewrite \eqref{e:split1} as
\begin{equation}\label{e:split2}
\rho_0 = \sum_{\alpha\in\Gamma}w_\alpha \tilde{A}_\alpha
\end{equation}
with
\begin{equation}
\tilde{A}_\alpha = \tfrac{1}{2}\left(A_\alpha + A_\alpha^\prime\right) = \tfrac{1}{2}(\id+\tilde{\mvec{r}}_\alpha\cdot\mvec{\sigma})
\end{equation}
and
\begin{subequations}
\begin{align}
\tilde{\mvec{r}}_{(0,0)}&=(1,0,1),\label{e:rt1}\\
\tilde{\mvec{r}}_{(0,1)}&=(-1,0,1),\\
\tilde{\mvec{r}}_{(1,0)}&=(1,0,-1),\\
\tilde{\mvec{r}}_{(1,1)}&=(-1,0,-1).\label{e:rt4}
\end{align}
\end{subequations}
While the new phase point operators $\tilde{A}_\alpha$ are equivalent to the old ones in being just simple linear combinations, the various choices are crucially different at later times due to the nonlinearity of the equations \eqref{e:1st_order_param}--\eqref{e:2nd_order_param} under which the classical spin vectors \eqref{e:rt1}--\eqref{e:rt4} evolve in time.

\begin{figure}\centering
\includegraphics[width=\linewidth]{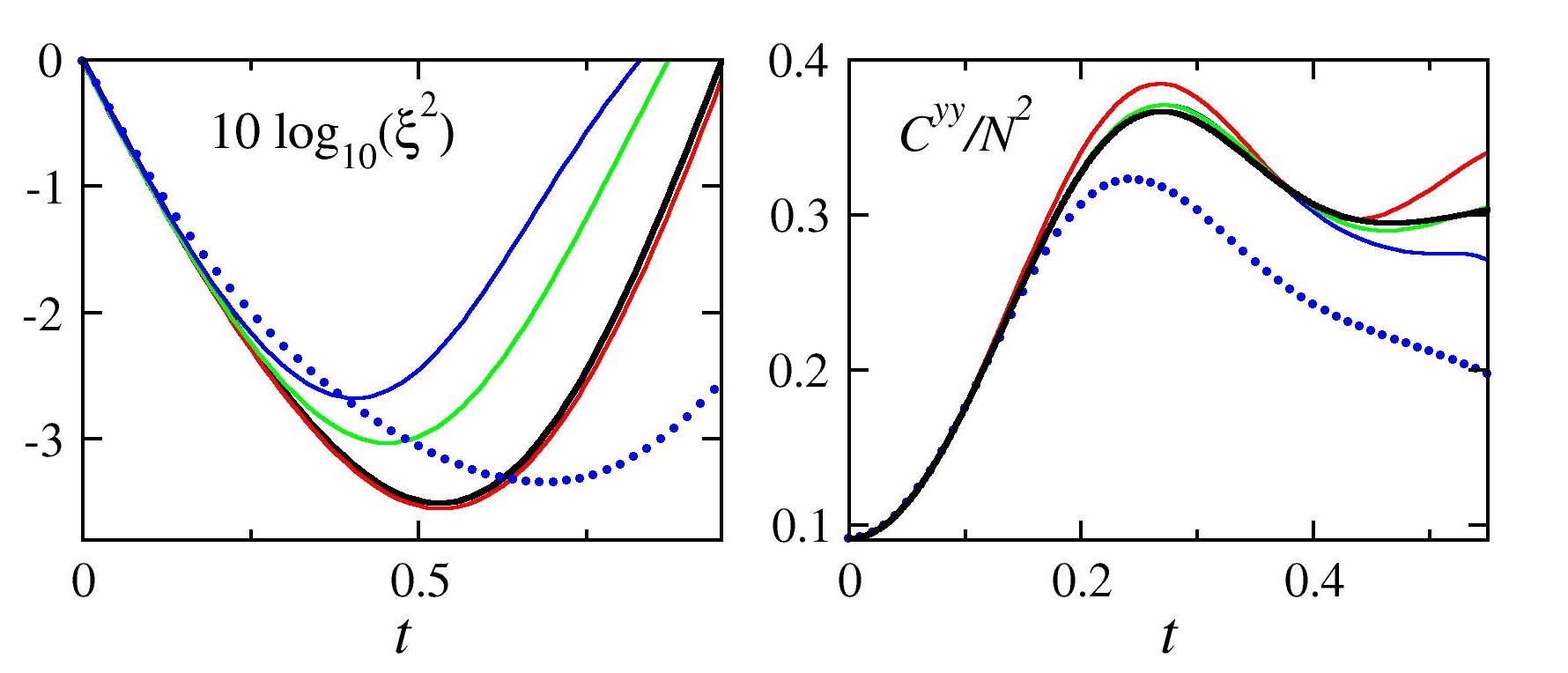}
\caption{\label{f:sampling}%
Comparison of the different sampling schemes $S_{\text{1-1}}$ (blue), $S_{\text{0-1}}$ (red), and $S_{\text{all}}$ (green) as defined in \eqref{e:S11}--\eqref{e:Sall}, and the method of SPR (blue dots). Sample sizes are $n=10^5$ in all cases. Left: Time evolution of the decibel spin squeezing for an $XX$ chain of 100 sites and long-range exponent $\alpha=3$. DMRG results are shown in black. Right: Total connected correlations $C^{yy}$ \eqref{e:Cmumu} of an Ising chain of 11 sites in a transverse field $\mvec{h}=(1,0,0)$ with long-range exponent $\alpha=1/2$. Results obtained by exact diagonalization are shown in black.
}%
\end{figure}

\begin{figure}\centering
\includegraphics[width=\linewidth]{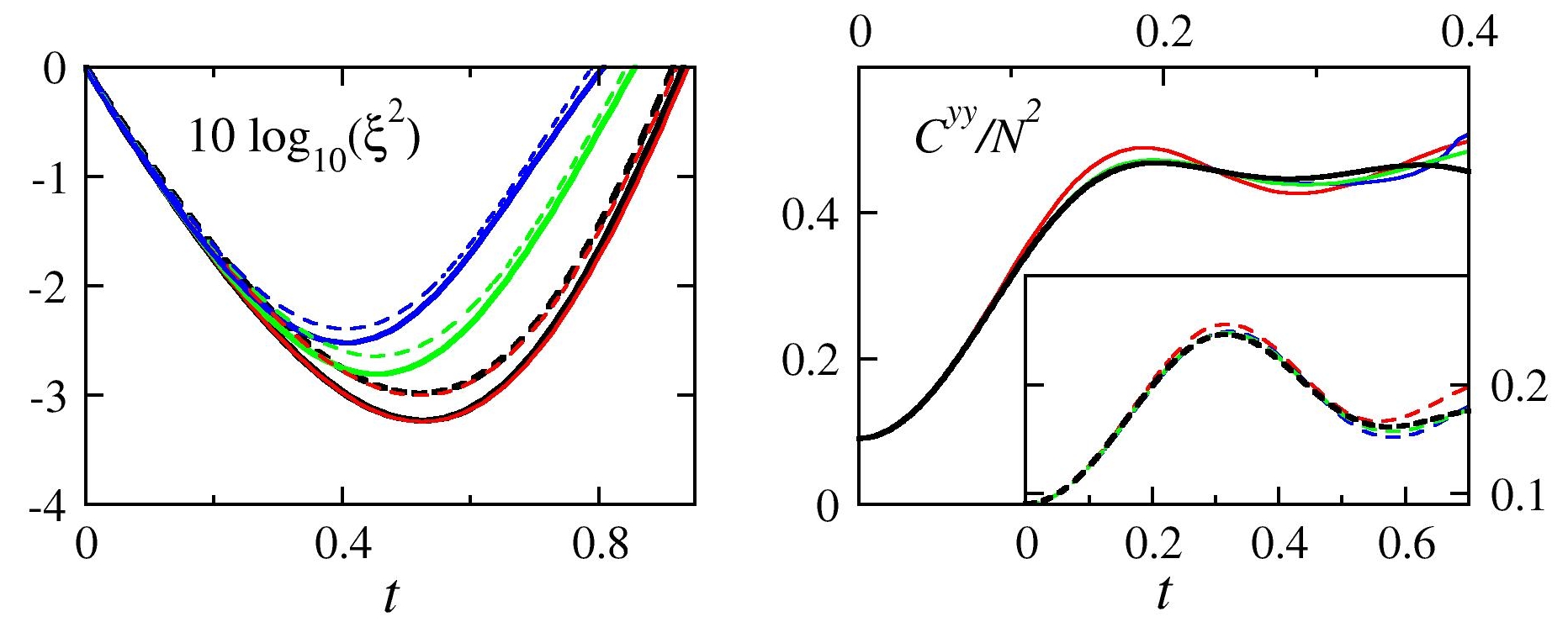}
\caption{\label{f:sampling_2}%
Comparison of the sampling schemes $S_{\text{1-1}}$ (blue), $S_{\text{0-1}}$ (red), and $S_{\text{all}}$ (green) as defined in \eqref{e:S11}--\eqref{e:Sall}. Sample sizes are $n=10^5$. Left: Time evolution of decibel spin squeezing for an $XX$ chain with long-range exponent $\alpha=3$ for a chain of 11 sites (dashed) and 20 sites (solid). For comparison, exact diagonalization results (black dashed) are shown for $N=11$, and DMRG results (black solid) for $N=20$. Right: Total connected correlations $C^{yy}$ \eqref{e:Cmumu} of an Ising chain of 11 sites in a transverse field $\mvec{h}=(1,0,0)$ with long-range exponent $\alpha=0$ (solid) and $\alpha=1$ (dashed). Results obtained by exact diagonalization are shown in black.
}%
\end{figure}

Out of the many possible choices of sampling schemes, we have experimented, always for $\rho_0 = (\id+\sigma^x)/2$, with the following sets of vectors to define the phase point operators,
\begin{subequations}
\begin{align}
S_{\text{1-1}}&=\bigl\{\mvec{r}_{(0,0)},\mvec{r}_{(1,0)},\mvec{r}_{(0,0)}^\prime,\mvec{r}_{(1,0)}^\prime\bigr\},\label{e:S11}\\
S_{\text{0-1}}&=\bigl\{\tilde{\mvec{r}}_{(0,0)},\tilde{\mvec{r}}_{(1,0)},\tilde{\mvec{r}}_{(0,0)}^\prime,\tilde{\mvec{r}}_{(1,0)}^\prime\bigr\},\\
S_{\text{all}}&=S_{\text{1-1}}\cup S_{\text{0-1}},\label{e:Sall}
\end{align}
\end{subequations}
where
\begin{equation}
\tilde{\mvec{r}}_{(0,0)}^\prime=(1,1,0),\qquad \tilde{\mvec{r}}_{(1,0)}^\prime=(1,-1,0).
\end{equation}
Numerical results obtained with $S_{\text{1-1}}$, $S_{\text{0-1}}$, and $S_{\text{all}}$ start to differ from each other after some time, as illustrated in Fig.~\ref{f:sampling}. To understand which of the sampling schemes performs better for a given problem, we analyze the accuracy of the results for different parameter values. As illustrated in Fig.~\ref{f:sampling_2}, it depends on the model, but only on the model, which of the sampling schemes performs better. For the $XX$ chain, we find that the $S_{\text{0-1}}$ sampling gives the best results for all system sizes studied. For the Ising chain in a transverse field, the $S_{\text{all}}$ sampling scheme performs best for various long-range exponents $\alpha$. This observation provides us with a method of how to choose a suitable sampling scheme for a given model. We can perform simulations for small system sizes where exact results for comparison are available, identify which sampling scheme works best for a given type of Hamiltonian, and then apply this 
optimal scheme to larger system sizes. For further improvement, however, a deeper theoretical understanding, and therefore being able to {\em a priori}\/ choose the optimal sampling for a given problem, would be desirable.

\section{B. Truncated time evolution in Pauli representation}
\setcounter{section}{2}
\setcounter{equation}{0}
\setcounter{figure}{0}

In this section the semi-classical time evolution equations for the Pauli expansion coefficients $a_i^\mu$ and $c_{ij}^{\mu\nu}$ are derived.

Starting from the Liouville-von Neumann equation \eqref{VNeqdWA} with the general Hamiltonian \eqref{e:Hgen} and taking partial traces $\Tr_{\not{\,i}}$, $\Tr_{\not{\,i}\not{\,j}}$, \dots, the BBGKY hierarchy of $N$ coupled differential equations is obtained,
\begin{subequations}
\begin{align}
\ii\partial_t \mathscr{A}_i=&\Com{H_i}{\mathscr{A}_i}+\sum_{k\neq i}\Tr\Com{H_{ik}}{\mathscr{A}_{ik}}\label{e:1st_order_app1}\\
\ii\partial_t \mathscr{A}_{ij}=&\Com{H_i+H_j+H_{ij}}{\mathscr{A}_{ij}}\nonumber\\
&+\sum_{k\neq i,j}\Tr_k\Com{H_{ik}+H_{jk}}{\mathscr{A}_{ijk}}
\label{e:2nd_order_app1}
\end{align}
\end{subequations}
Inserting the cluster expansion \eqref{e:cluster1} and \eqref{e:cluster2} and rearranging terms, the hierarchy takes the form 
\begin{subequations}
\begin{align}
\ii\partial_t \mathscr{A}_i=&\Com{H_i}{\mathscr{A}_i}+\sum_{k\neq i}\Tr\Com{H_{ik}}{\mathscr{C}_{ik}+\mathscr{A}_i \mathscr{A}_k}\label{e:1st_order_app2}\\
\ii\partial_t \mathscr{C}_{ij}=&\Com{H_i+H_j+H_{i\not{\,j}}^\text{H}+H_{j\not{\,i}}^\text{H}}{\mathscr{C}_{ij}}+\Com{H_{ij}}{\mathscr{C}_{ij}+\mathscr{A}_i \mathscr{A}_j}\nonumber\\
&+\sum_{k\neq i,j}\left(\Tr_k\Com{H_{ik}}{\mathscr{A}_i \mathscr{C}_{jk}}+\Tr_k\Com{H_{jk}}{\mathscr{A}_j \mathscr{C}_{ik}}\right)\nonumber\\
&-\mathscr{A}_i\Tr_i\Com{H_{ij}}{\mathscr{C}_{ij}+\mathscr{A}_i \mathscr{A}_j}\nonumber\\
&-\mathscr{A}_j\Tr_j\Com{H_{ij}}{\mathscr{C}_{ij}+\mathscr{A}_i \mathscr{A}_j}\nonumber\\
&+\sum_{k\neq i,j}\left(\Tr_k\Com{H_{ik}}{\mathscr{C}_{ijk}}+\Tr_k\Com{H_{jk}}{\mathscr{C}_{ijk}}\right)
\label{e:2nd_order_app2}
\end{align}
\end{subequations}
Neglecting the 3-spin correlations $\mathscr{C}_{ijk}$, the first two equations of the hierarchy decouple from the rest, as in Eqs.~\eqref{e:1st_order} and \eqref{e:2nd_order} of the main text. Next we use the Pauli representations \eqref{e:Aexp} and \eqref{e:Cexp} of the $\mathscr{A}$- and $\mathscr{C}$-operators, and also expand the Hamiltonian \eqref{e:Hgen} in that basis,
\begin{equation}
H=-\sum_{\mu\in\{x,y,z\}}\sum_{i\neq j}J_{ij}^\mu\sigma_i^\mu\sigma_j^\mu - \mvec{h}\cdot\sum_i\mvec{\sigma}_i
\end{equation}
Inserting the representations into the truncated BBGKY equations \eqref{e:1st_order} and \eqref{e:2nd_order}, one makes use of the orthogonality of the Pauli operators to calculate, by methods similar to the Appendix of Ref.~\cite{PaskauskasKastner12}, the following time evolution equations for the expansion coefficients $a_i^\mu$ and $c_{ij}^{\mu\nu}$ for all $i,j=1,\dotsc,N$ and $\mu,\nu\in\{x,y,z\}$,
\begin{subequations}
\begin{equation}
 \frac{1}{2}\partial_t a_i^\mu=-\sum_{\beta\gamma}\biggl[\biggl(h^\beta+\sum_{k\neq i}J_{ik}^\beta a_k^\beta\biggr)a_i^\gamma+\sum_{k\neq i}J_{ik}^\beta c_{ki}^{\beta\gamma}\biggr]\varepsilon^{\mu \beta\gamma},\label{e:1st_order_param}
\end{equation}
\begin{widetext}
\begin{multline}
\frac{1}{2}\partial_t c_{ij}^{\mu\nu}=-\sum_{\beta}(J_{ij}^\nu a_i^\beta-J_{ij}^\mu a_j^\beta)\varepsilon^{\mu\nu\beta}-\sum_{\beta,\gamma}\biggl[\biggl(h^\beta+\sum_{k\neq i,j}J_{ik}^\beta a_k^\beta\biggr)c_{ij}^{\gamma\nu}\varepsilon^{\beta \gamma\mu}+\biggl(h^\beta+\sum_{k\neq i,j}J_{jk}^\beta a_k^\beta\biggr) c_{ij}^{\mu \gamma}\varepsilon^{\beta \gamma\nu}\biggr]\\
 -\sum_{\beta,\gamma}\sum_{k\neq i,j}\left[J_{ik}^\beta a_i^\gamma c_{jk}^{\nu\beta}\varepsilon^{\beta \gamma\mu}+J_{jk}^\beta a_j^\gamma c_{ik}^{\mu\beta}\varepsilon^{\beta \gamma\nu}\right]+\sum_{\beta,\gamma}J_{ij}^\beta \left[a_i^\mu(c_{ij}^{\beta \gamma}+a_i^\beta a_j^\gamma)\varepsilon^{\beta \gamma \nu}+a_j^\nu(c_{ij}^{\gamma\beta}+a_i^\gamma a_j^\beta)\varepsilon^{\beta \gamma \mu}\right].
\label{e:2nd_order_param}
\end{multline}
\end{widetext}
\end{subequations}

\section{C. Ambiguities in the calculation of expectation values}
\setcounter{section}{3}
\setcounter{equation}{0}
\setcounter{figure}{0}

Due to the approximations made in the method described in this Letter (and also the one of SPR \cite{Schachenmayer_etal15}), ambiguities may arise when calculating expectation values. This may occur when approximating the expectation value of an operator written in two equivalent ways, like in $\sigma^z=\ii\sigma^x\sigma^y$. Calculating expectation values along the lines of \eqref{e:1spin}, one could compute either
\begin{subequations}
\begin{equation}
\left\langle\sigma_i^z\right\rangle =\sum_{\mvec{\alpha}\in\Gamma^N}w_{\alpha_1}\cdots w_{\alpha_N} a_i^z \approx \sum_{\mvec{\alpha}\in S_n}a_i^z,
\end{equation}
or
\begin{equation}
\left\langle\sigma_i^z\right\rangle =\ii\sum_{\mvec{\alpha}\in\Gamma^N}w_{\alpha_1}\cdots w_{\alpha_N} a_i^x a_i^y \approx \ii\sum_{\mvec{\alpha}\in S_n}a_i^x a_i^y.
\end{equation}
\end{subequations}
The latter does not seem to be a reasonable choice though, as the expectation value of a Hermitian operator calculated this way turns out to be imaginary, and it appears favorable to express an operator in the simplest possible way.

When calculating total correlations as in Figs.~\ref{f:TFIM}--\ref{f:2d}, a similar kind of ambiguity arises. Calculating expectation values according to \eqref{e:1spin} and \eqref{e:2spin}, one may either write
\begin{subequations}
\begin{equation}\label{e:C_SPR}
\left\langle\sum_{i,j}\sigma_i^\mu\sigma_j^\mu\right\rangle
\approx \frac{1}{n}\sum_{\mvec{\alpha}\in S_n}\sum_{i,j}\left(a_i^\mu a_j^\mu + c_{ij}^{\mu\mu}\right)
\end{equation}
where $c_{ii}^{\mu\mu}\equiv0$, or 
\begin{equation}\label{e:C_PRK}
N+\left\langle\sum_{i\neq j}\sigma_i^\mu\sigma_j^\mu\right\rangle
\approx N+\frac{1}{n}\sum_{\mvec{\alpha}\in S_n}\sum_{i\neq j}\left(a_i^\mu a_j^\mu + c_{ij}^{\mu\mu}\right)
\end{equation}
\end{subequations}
While the left-hand sides of \eqref{e:C_SPR} and \eqref{e:C_PRK} are identical, the approximations on the right-hand sides differ in general. The choice in \eqref{e:C_SPR} amounts to approximating, for the terms where $i=j$, the expectation value of the identity operator by a value different from 1. We have opted for \eqref{e:C_PRK} in our method, and we have verified that this choice gives better results. Schachenmayer {\em et al.}\ in \cite{Schachenmayer_etal15} used \eqref{e:C_SPR}. We have tested that, in their method, independently of whether \eqref{e:C_SPR} or \eqref{e:C_PRK} is used, deviations from the exact result become visible at around similar times. Surprisingly, for the method of SPR, qualitative agreement between exact and approximate results can be better with \eqref{e:C_SPR}, and we used this definition when comparing their method to ours. Since our method yields more accurate results than SPR in all cases, we advocate the use of \eqref{e:C_PRK} in general.

\section{D. Exact Ising results}
\setcounter{section}{4}
\setcounter{equation}{0}
\setcounter{figure}{0}

For the quantum Ising model with a longitudinal magnetic field and arbitrary couplings $J_{ij}$,
\begin{equation}
H=-\sum_{i\neq j}J_{ij}\sigma_i^z\sigma_j^z - h\sum_i\sigma_i^z,
\end{equation}
we show that the middle equation in \eqref{e:2spin}, evaluated by making use of the truncated time evolution equations \eqref{e:1st_order_param}--\eqref{e:2nd_order_param}, agrees with the exact analytic solution of the Ising model. As a consequence, the estimate on the right-hand side of \eqref{e:2spin} becomes exact in the limit of large sample size.
For simplicity we set $h=0$ in the following, but an analogous calculation can be performed for nonzero magnetic field, recovering the exact solution also in this case.

When using the sampling $S_{\text{1-1}}$ defined in \eqref{e:S11}, the system of equations \eqref{e:1st_order_param}--\eqref{e:2nd_order_param} reduces to
\begin{subequations}
\begin{equation}
 \partial_t a_i^x=2a_i^y\beta_i,\qquad
 \partial_t a_i^y=-2a_i^x\beta_i
\end{equation}
and
\begin{equation}\label{corr_Ising}
\begin{split}
 \partial_t c_{ij}^{xy}=&2\beta_{i\not{\,j}}c_{ij}^{yy}-2\beta_{j\not{\,i}}\,c_{ij}^{xx}-2J_{ij}(a_j^za_i^ya_j^y-a_i^za_i^xa_j^x)\\
 \partial_t c_{ij}^{yx}=&2\beta_{j\not{\,i}}\,c_{ij}^{yy}-2\beta_{i\not{\,j}}c_{ij}^{xx}-2J_{ij}(a_i^za_i^ya_j^y-a_j^za_i^xa_j^x)\\
 \partial_t c_{ij}^{xx}=&2\beta_{i\not{\,j}}c_{ij}^{yx}+2\beta_{j\not{\,i}}\,c_{ij}^{xy}-2J_{ij}(a_j^za_i^ya_j^x+a_i^za_i^xa_j^y)\\
 \partial_t c_{ij}^{yy}=&-2\beta_{i\not{\,j}}c_{ij}^{xy}-2\beta_{j\not{\,i}}\,c_{ij}^{yx}+2J_{ij}(a_j^za_i^ya_j^x+a_i^za_i^ya_j^x)
\end{split}
\end{equation}
\end{subequations}
where we introduced the constants
\begin{equation}
\beta_{i}=\sum_{k\neq i} J_{ik} a_k^z(0),\qquad \beta_{i\not{\,j}}=\sum_{k\neq i,j} J_{ik} a_k^z(0).
\end{equation}
The $a_i^z$ do not change in time, and $c_{ij}^{xz}$, $c_{ij}^{yz}$, and $c_{ij}^{zz}$ remain zero for all times.
The quantities $a_i^x$ and $a_i^y$ evolve independently from all $c_{ij}$, 
\begin{subequations}
\begin{align}
a_i^x(t)=&a_i^x(0)\cos(2t\beta_i)+a_i^y(0)\sin(2t\beta_i),\\
a_i^y(t)=&a_i^y(0)\cos(2t\beta_i)-a_i^x(0)\sin(2t\beta_i),
\end{align}
\end{subequations}
where $a_i^x(0)=1$. The equations in \eqref{corr_Ising} are local in the sense that they relate only quantities with the same indices $i$ and $j$. We can rewrite these equations in matrix form,
\begin{subequations}
\begin{equation}
 \partial_t\mvec{c_{ij}}=M_{ij}\mvec{c}_{ij}+\mvec{f}_{ij}(t)\\
\end{equation}
\pagebreak
with
\begin{align}
 \mvec{c}_{ij}^T&=\left(c_{ij}^{xy},c_{ij}^{yx},c_{ij}^{xx},c_{ij}^{yy}\right),\\
 M_{ij}&=2\begin{pmatrix}
   0 & 0 & -\beta_{j\not{\,i}} & \beta_{i\not{\,j}}\\
   0 & 0 & -\beta_{i\not{\,j}} & \beta_{j\not{\,i}}\\
   \beta_{j\not{\,i}} & \beta_{i\not{\,j}} & 0 & 0 \\
   -\beta_{i\not{\,j}} & -\beta_{j\not{\,i}} & 0 & 0 
  \end{pmatrix},
\end{align}
\begin{equation}
 \mvec{f}_{ij}(t)=2J_{ij}\begin{pmatrix}
                      a_i^z(0)a_i^x(t)a_j^x(t)-a_j^z(0)a_i^y(t)a_j^y(t)\\
                      -a_i^z(0)a_i^y(t)a_j^y(t)+a_j^z(0)a_i^x(t)a_j^x(t)\\
                      -a_i^z(0)a_i^x(t)a_j^y(t)-a_j^z(0)a_i^y(t)a_j^x(t)\\
                      a_i^z(0)a_i^y(t)a_j^x(t)+a_j^z(0)a_i^y(t)a_j^x(t)
                      \end{pmatrix}.
\end{equation}
\end{subequations}
Expressing $\mvec{f_{ij}}$ in terms of the four eigenvectors
\begin{subequations}
\begin{align}
 \mvec{v_1}=\begin{pmatrix}
                        \ii\\
                        \ii\\
                        -1\\
                        1
                       \end{pmatrix},\;
 \mvec{v_2}=\begin{pmatrix}
                        -\ii\\
                        \ii\\
                        1\\
                        1
                       \end{pmatrix},\;
 \mvec{v_3}=(\mvec{v_{ij}^2})^*,\;
 \mvec{v_4}=(\mvec{v_{ij}^1})^*
\end{align}
of $M$ with corresponding eigenvalues $\lambda_{ij}^1$, $\lambda_{ij}^2$, $(\lambda_{ij}^2)^*$, and $(\lambda_{ij}^1)^*$,
where
\begin{equation}
\lambda_{ij}^1=-2\ii (\beta_{i\not{\,j}}+\beta_{j\not{\,i}}),\qquad \lambda_{ij}^2=2\ii (\beta_{i\not{\,j}}-\beta_{j\not{\,i}}),
\end{equation}
\end{subequations}
we obtain the solution
\begin{widetext}
\begin{equation}
\begin{split}
 \mvec{c_{ij}}(t)=&\int_0^t \dd\tau\, \ee^{M(t-\tau)}\mvec{f_{ij}}(\tau)\\
=&\frac{\ii}{4}\left[\ee^{-2\ii (\beta_{i\not{\,j}}+\beta_{j\not{\,i}})t}\ee^{-2\ii J_{ij}(a_i^z+a_j^z)t}-\ee^{-2\ii (\beta_{i\not{\,j}}+\beta_{j\not{\,i}})t}\right]\left[\ii (a_i^y(0)a_j^y(0)-1)+(a_i^y(0)+a_j^y(0))\right]\mvec{v_1}\\
 &+\frac{\ii}{4} \left[\ee^{2\ii (\beta_{i\not{\,j}}-\beta_{j\not{\,i}})t}\ee^{2\ii J_{ij}(a_j^z-a_i^z)t}-\ee^{2\ii (\beta_{i\not{\,j}}-\beta_{j\not{\,i}})t}\right]\left[\ii (a_i^y(0)a_j^y(0)+1)+(a_i^y(0)-a_j^y(0))\right]\mvec{v_2}+\text{c.c.}
\end{split}
\end{equation}
Next we compute the correlation functions according to the right-hand side of \eqref{e:2spin}, i.e., from an infinite sample of initial states drawn according to the sampling scheme $S_{\text{1-1}}$ defined in \eqref{e:S11}. As a first step, we sum only over the possible initial values, as given by that sampling, of the $a(0)$ components at lattice sites $i$ and $j$, yielding
\begin{subequations}
\begin{align}
 &\frac{1}{16}\sum_{a_i^z,a_i^y,a_j^z,a_j^y\in S_{\text{1-1}}}\mvec{c_{ij}}(t)=
 \frac{1}{2}\left[\cos^2(2J_{ij}t)-1\right]\begin{pmatrix}
        \sin\left[2(\beta_{i\not{\,j}}+\beta_{j\not{\,i}})t\right]-\sin\left[2(\beta_{i\not{\,j}}-\beta_{j\not{\,i}})t\right]\\
        \sin\left[2(\beta_{i\not{\,j}}+\beta_{j\not{\,i}})t\right]+\sin\left[2(\beta_{i\not{\,j}}-\beta_{j\not{\,i}})t\right]\\
        -\cos\left[2(\beta_{i\not{\,j}}+\beta_{j\not{\,i}})t\right]-\cos\left[2(\beta_{i\not{\,j}}-\beta_{j\not{\,i}})t\right]\\
        \cos\left[2(\beta_{i\not{\,j}}+\beta_{j\not{\,i}})t\right]-\cos\left[2(\beta_{i\not{\,j}}-\beta_{j\not{\,i}})t\right]
       \end{pmatrix},\\
 &\frac{1}{16}\sum_{a_i^z,a_i^y,a_j^z,a_j^y\in S_{\text{1-1}}}
 \begin{pmatrix}
 a_i^x(t)a_j^y(t)\\
 a_i^y(t)a_j^x(t)\\
 a_i^x(t)a_j^x(t)\\
 a_i^y(t)a_j^y(t)
 \end{pmatrix}
=
 -\frac{1}{2}\cos^2(2J_{ij}t)\begin{pmatrix}
        \sin\left[2(\beta_{i\not{\,j}}+\beta_{j\not{\,i}})t\right]-\sin\left[2(\beta_{i\not{\,j}}-\beta_{j\not{\,i}})t\right]\\
        \sin\left[2(\beta_{i\not{\,j}}+\beta_{j\not{\,i}})t\right]+\sin\left[2(\beta_{i\not{\,j}}-\beta_{j\not{\,i}})t\right]\\
        -\cos\left[2(\beta_{i\not{\,j}}+\beta_{j\not{\,i}})t\right]-\cos\left[2(\beta_{i\not{\,j}}-\beta_{j\not{\,i}})t\right]\\
        \cos\left[2(\beta_{i\not{\,j}}+\beta_{j\not{\,i}})t\right]-\cos\left[2(\beta_{i\not{\,j}}-\beta_{j\not{\,i}})t\right]
       \end{pmatrix}.
\end{align}
\end{subequations}
Here we identified $a_i^\mu\equiv a_i^\mu(0)$, i.e., $a$-coefficients without a time argument denote initial values. (Note that the coefficients $\beta_{i\not{\,j}}$ and $\beta_{j\not{\,i}}$ depend on $a_k^z$.) The terms proportional to $\cos^2(2J_{ij}t)$ cancel out and the final result is obtained by summing over the remaining $a_k^z$ with $k\neq i,j$,
\begin{align}
\begin{pmatrix}
        \left\langle\sigma_i^x\sigma_j^y\right\rangle(t)\\
        \left\langle\sigma_i^y\sigma_j^x\right\rangle(t)\\
        \left\langle\sigma_i^x\sigma_j^x\right\rangle(t)\\
        \left\langle\sigma_i^y\sigma_j^y\right\rangle(t)
       \end{pmatrix}    
 =\frac{2}{4^{N-1}}\sum_{\genfrac{}{}{0pt}{}{\scriptstyle a_k^z,a_k^y\in S_{\text{1-1}}}{\scriptstyle k\neq i,j}}
\begin{pmatrix}
        -\sin\left[2(\beta_{i\not{\,j}}+\beta_{j\not{\,i}})t\right]+\sin\left[2(\beta_{i\not{\,j}}-\beta_{j\not{\,i}})t\right]\\
        -\sin\left[2(\beta_{i\not{\,j}}+\beta_{j\not{\,i}})t\right]-\sin\left[2(\beta_{i\not{\,j}}-\beta_{j\not{\,i}})t\right]\\
        \cos\left[2(\beta_{i\not{\,j}}+\beta_{j\not{\,i}})t\right]+\cos\left[2(\beta_{i\not{\,j}}-\beta_{j\not{\,i}})t\right]\\
        -\cos\left[2(\beta_{i\not{\,j}}+\beta_{j\not{\,i}})t\right]+\cos\left[2(\beta_{i\not{\,j}}-\beta_{j\not{\,i}})t\right]
       \end{pmatrix}
 =
 \begin{pmatrix}
        0\\
        0\\
        P_{ij}^-+P_{ij}^+\\
        P_{ij}^--P_{ij}^+
       \end{pmatrix}
\end{align}
\end{widetext}
with
\begin{equation}
P_{ij}^\pm=\frac{1}{2}\prod_{k\neq i,j}\cos\left[2(J_{ki}\pm J_{kj})t\right],
\end{equation}
which agrees with the exact analytic result found in \cite{vdWorm_etal13}. In summary, for the Ising model we recover from the truncated time evolution equations \eqref{e:1st_order_param}--\eqref{e:2nd_order_param} the exact analytic solutions for all one-point and two-point functions. This is a significant improvement compared to the method in \cite{Schachenmayer_etal15}.


\end{document}